# Eukaryotic gene regulation at equilibrium, or non?

Benjamin Zoller (1,2,3), Thomas Gregor (1,2,3), Gašper Tkačik (4)

1 Lewis-Sigler Institute for Integrative Genomics, Princeton University, Princeton, NJ, USA
2 Joseph Henry Laboratories of Physics, Princeton University, Princeton NJ, USA
3 Department of Developmental and Stem Cell Biology UMR3738, Institut Pasteur, Paris, France
4 Institute of Science and Technology Austria, Klosterneuburg, Austria

## Abstract
Models of transcriptional regulation that assume equilibrium binding of transcription factors have been very successful at predicting gene expression from sequence in bacteria. However, analogous equilibrium models do not perform as well in eukaryotes, most likely due to the largely out-of-equilibrium nature of eukaryotic regulatory processes. These processes come with unavoidable energy expenditure at the molecular level to support precise, reliable, or fast gene expression responses that could correspond to evolutionarily optimized regulatory strategies. Unfortunately, the space of possible non-equilibrium mechanisms is vast and predominantly uninteresting. The key question is therefore how this space can be navigated efficiently, to focus on mechanisms and models that are biologically relevant. In this review, we advocate for the normative role of theory – theory that prescribes rather than just describes – in providing such a focus. Theory should expand its remit beyond inferring models from data (by fitting), towards identifying non-equilibrium gene regulatory schemes (by optimizing) that may have been evolutionarily selected due to their favorable functional characteristics which outperform regulation at equilibrium. This approach can help us navigate the expanding complexity of regulatory architectures, and refocus the questions about the observed biological mechanisms from *how* they work towards *why* they have evolved in the first place. We illustrate our reasoning by toy examples for which we provide simulation code.

## Non-equilibrium processes essential to life consume energy

A defining physical property of life is its non-equilibrium nature (a cell at thermodynamic equilibrium with its surrounding is typically a dead cell). Indeed, living organisms are characterized as thermodynamically open systems involving the exchange of heat, work and matter with the environment, in order to achieve self-organization, growth, homeostasis, adaptation, and so on. The non-equilibrium activity of cells is typically driven by internal enzymatic processes that burn ATP rather than by external forces [1].

In addition to the exchange of matter, cells also carefully orchestrate flows of information, by tying their internal chemical state to changes in the environment [2]. Typical examples include the chemotaxis network of *E. coli*, where information about chemoattractant gradient is detected by receptors at the cell surface and transduced to the motors driving the flagella [3], and MAPK signaling pathways in yeast, where information propagates through phosphorylation in a cascade of signaling molecules before affecting gene expression [4]. Chemical modifications at the heart of these pathways consume ATP and are thus intrinsically kept far from equilibrium. More generally, information processing in biological systems can be energetically costly, especially when speed and precision are at a premium, as substantiated by well-known examples in neuroscience [5,6].

In contrast, gene regulation has historically been studied through the lens of chemical processes at equilibrium, which are greatly successful in prokaryotes. There, the "thermodynamic model of gene regulation" assumes that the equilibrium occupancy of transcription factors (TFs) on their regulatory sites on the DNA dictates downstream gene expression, and this provides a dramatic conceptual and predictive simplification of the regulatory processes [7]. With some notable exceptions [8], thermodynamic models in bacteria have enabled a concordance between *in vivo* and *in vitro* as well as between steady-state and kinetic measurements [9,10].

But why should nature limit itself to equilibrium regulatory processes? We stress that what we denominate here as equilibrium (EQ) versus non-equilibrium (NEQ) depends on the boundaries of the system under consideration. It is clear that transcription as a whole is a NEQ process, as energy is expended by the enzyme RNA polymerase (RNAP) to synthetize the RNA polymer. What we are focusing on here are the regulatory steps preceding RNAP activity. These steps involve the TFs that govern the likelihood of RNAP binding and unbind and thus *control* productive transcription. With this clarification in mind, the key argument for regulation at equilibrium is either one of modeling simplicity (Ockham's razor, time-scale separation, coarse-graining) [10] or one of energy expenditure minimization [3] – which could be evolutionary selected for.

Given our current estimates, however, the energetic costs of transcriptional regulation should be negligible compared to other known energy-consuming cellular processes. The average cell in the human body can produce between $10^8$ - $10^9$ ATP/s, of which motility consumes $10^5$ - $10^6$ ATP/s and protein production more than $10^7$ ATP/s (by maintaining typical protein concentrations) [11]. The overall metabolic cost to maintain a eukaryotic cell is around $10^7$ ATP/s, while making a new daughter cell costs around $10^{12}$ ATP in total. Overall, the cost of DNA replication is one to two orders of magnitude smaller than the cost of transcription (synthesis and degradation), which is a further one to two orders of magnitude smaller than the cost of translation (synthesis and degradation) [12]. The amount of ATP needed to support the actin cytoskeleton has been estimated to be around 50% of the total ATP consumption of a cell [13]. Thus, the vast majority of energy consumed by a cell appears dedicated to protein renewal and cytoskeleton rearrangement. Taken together, it is difficult to imagine how any putative ATP costs of regulation and initiation (which are most likely smaller than the follow-up cost of productive transcription and translation) could constitute a relevant dent in the cell's total energy budget. We should therefore turn our attention to other costs and benefits of non-equilibrium regulatory schemes.

In the following sections we briefly review the basic NEQ mechanisms and their possible regulatory benefits. Then we give an overview of current evidence for NEQ regulation in eukaryotes, and highlight the necessity care required to properly assess which signatures support or rule out NEQ and EQ mechanisms. By means of toy models we demonstrate that the space of NEQ regulatory models is vast, but only a fraction of this space leads to functionally relevant regulatory phenotypes. Lastly, we propose one particular option to navigate the complexity inherent to NEQ models, by postulating that regulatory mechanisms evolved to optimize some of these regulatory phenotypes. This normative approach can generate both functional (what has been selected) and mechanistic (how it works) hypotheses, thus bringing together aspects of evolutionary and molecular biology.

## Fundamentals of non-equilibrium regulatory processes

To highlight the possible regulatory benefits of NEQ mechanisms, we adopt a biophysically-rooted kinetic description of regulatory processes. Within that framework, equilibrium has a very specific meaning, namely the thermodynamic equilibrium of the underlying molecular processes, which can be traced back to their time-reversibility. These qualifications are important. First, it is possible to consider truly NEQ (kinetic) processes but coarse-grain them such that their effective description resembles EQ [10,14,15]; this is not our focus here. Second, it is important to differentiate between free energy dissipation during a transient relaxation to the steady state which can be discussed also for EQ systems, and dissipation in the true NEQ steady state which is unique to NEQ systems; here, we will mainly discuss the latter. In what follows, we employ a general "linear framework" based on the linear chemical master equation. It describes the relevant processes in terms of chemical reaction networks [16–18] and defines their degree of irreversibility (i.e. their non-equilibrium character) through the violation of detailed balance.

Most reaction networks are NEQ. Indeed, barring very particular network topologies (such as sequences or trees), a random assignment of reaction rates between the vertices of the network will inevitably yield a NEQ system. This is because EQ models must implement strict relationships between their reactions rates to satisfy detailed balance, encoding for the reversibility of elementary processes [1]. By breaking detailed balance, one can access the full reaction network space, and, in particular, irreversible reaction cycles, the simplest NEQ structures which imply energy expenditure and dissipation [16,17].

Cycles have been widely used to describe molecular motors, phosphorylation cycles (cell cycle, circadian cycle) [19], transition state cycle of enzymes, and regulation by gene promoters (promoter progression) [20,21] (Fig. 1A-B). In particular, for transcriptional regulation, cycles have been shown to be sufficiently rich to implement a vast range of logical computations [22]. NEQ reaction networks also permit regulatory mechanisms with improved or novel functionality (see below). They have been the focus of several recent publications: improved fidelity and specificity (via proof-reading mechanisms) [23–25], gene expression noise reduction [26], improved information transmission [27], improved sensitivity (ultra-sensitivity, high Hill coefficients of the regulatory functions) [28,29], improved sensing (beyond Berg-Purcell limit) [30], faster relaxation [31,32], and improved timing precision [33,34].

**[Figure 1]** Reaction cycles as the fundamental "unit" of NEQ models highlight key signatures of irreversibility. **(A)** Three biological examples of irreversible cycles. (Left) A walking myosin on an actin filament. (Middle) An enzymatic cycle. (Right) Gene promoter progression towards activation. **(B)** A class of models that can continuously transform from EQ to NEQ is parametrized by a unique "reversibility" parameter $\alpha$ for a given size (N = 3 states), at constant mean state occupancies ($P_i = 1/N$) and residence times ($T_i = T_1 = 1$min). **(C)** Two stochastic realizations, for a fully reversible ($\alpha = 0$, left) and fully irreversible cycle ($\alpha = 1$, right). (Top) Individual state occupancies. (Bottom) Winding number for each realization, i.e. the number of counter-clockwise cycle completions. Irreversibility leads to a clear temporal ordering of reactions, as highlighted by the progression of the winding number. **(D)** Entropy production $\Sigma$ or dissipation ($\Sigma$ times temperature) as a function of the current J along the cycle. Both J and $\Sigma$ monotonically increase as the cycle approaches full irreversibility ($\alpha = 1$). The period of the cycle is related to the current as $1/J$, and the current is maximal, $J_{max} = 1 / (NT_1)$, when $\alpha = 1$. The entropy production tends to infinity when the cycle is fully irreversible. The presence of currents and entropy production are hallmarks of

NEQ reaction schemes. **(E)** Residence time distribution P(T$_{2->3}$) for the combined states 2 and 3, i.e., time spent in 2 and/or 3 before ending in 1. T$_{2->3}$ is phase-type distributed and its shape depends on α, changing from exponential-like (α = 0) to "peaked" (α = 1). Peaked residence time distributions are another signature of strongly irreversible processes. (Inset) residence time distribution P(T$_1$) for individual states. T$_1$ is exponentially distributed and does not depend on α.

### Direct experimental signatures of non-equilibrium regulation

The two most unequivocal signatures of NEQ processes at steady state are heat dissipation with entropy production and the irreversibility of the underlying reactions (Fig. 1B & D). These two features are intimately linked, since at least in principle knowing the amount of dissipated heat is informative about the degree of reversibility and vice versa [16,35]. Estimating energy expenditure in absolute units from chemical reaction models is a priori possible, but very challenging to apply in practice. As a first complication, one must know the exact degrees of freedom (the microstates) of the system to connect dissipation with kinetics. Missing microscopic details will bias the estimation since coarse-graining of the reaction network reduces the apparent dissipation [15]. Moreover, defining an adequate temperature for a molecular non-equilibrium system can be non-trivial.

Recent theoretical progress circumvents some of these issues, e.g., by quantifying fluxes in configurational phase space [36], by quantifying the irreversibility of fluctuations in timing of events from time series using semi-Markov processes [37], and by bounding the entropy production from time series using the waiting time statistics of hidden-Markov process [38]. Despite these theoretical advances, measuring heat dissipation in practice remains challenging and has been typically limited to whole organisms in the context of metabolism [39,40], or to *in vitro* molecular systems of limited complexity, such as molecular motors and enzymatic reactions [41]. Such measurements for complex *in vivo* processes such as gene regulation are currently out of reach.

Irreversibility, however, can have consequences other than heat dissipation, which can be inferred from time series measurements of observables even in complex *in vivo* systems (Fig. 1C). For example, provided that the states of the system can be determined from data (such as configurations of the molecular motor, gene activity, etc.), one can estimate the distribution of residence times (times spent in a set of states, also known as dwell or waiting times), whose shape depends on the reversibility of the underlying reactions (Fig. 1E). It has been demonstrated that peaked, gamma-like distributions are a strong signature of irreversibility, and thus of NEQ processes [28,42].

Notably, the silent intervals between multiple rounds of transcriptional bursts in eukaryotic genes are well described by such peaked distributions [43,44], suggesting that the inactive periods of these genes arise from multiple irreversible sequential steps required before transcriptional initiation. Such a succession of irreversible steps leads to a refractory period [20,45], a minimal duration preventing immediate reactivation of transcription. The nature of these refractory periods and the underlying sequential steps remain poorly understood, and likely depend on the induction pathway. The steps could be related to sequential modifications of the chromatin template mediated by various transcription and pioneer factors, the establishment of contact between regulatory elements and promoter through chromatin looping, and the resulting formation of cofactor condensates leading to the assembly of the pre-initiation complex. New evidence suggests that transcription initiation stimulates cofactor condensates, whereas bursts stimulate dissolution, implementing a

NEQ feedback [46]. Such a negative feedback should typically lead to pulsatile bursts of transcription whose silent interval distributions are again peaked rather than exponential [47].

Regulatory phenotypes bearing indirect signatures of non-equilibrium regulation

Other, even more circumstantial signatures of NEQ regulation can be derived from response properties of enhancers or promoters, i.e., from their "regulatory phenotypes". For example, one can focus on gene regulatory functions (also known as input-output relations or induction curves, showing the gene expression level as a function of inducer or regulator concentration), and in particular on their specificity or sensitivity (Fig. 2) [29,48]. One can argue that NEQ schemes make it possible to achieve a certain high level of specificity or sensitivity, and in this context, recent measurements in eukaryotes tend to favor NEQ regulation schemes [49,50]. Such observations, however, do not constitute a definite proof. This is because even if current evidence is consistent with a proposed NEQ scheme, it typically does not rule out the existence of alternative EQ mechanisms which would be equally consistent with measured regulatory phenotypes.

We illustrate this discussion by building toy models that achieve two desired regulatory phenotypes, one for high specificity (Fig. 2A, left) and another for high sensitivity (Fig. 2A, right). The high-specificity model achieves good discrimination between specific and non-specific TF activation by using a NEQ kinetic proof-reading scheme, thereby outperforming alternatives at equilibrium (Fig. 2B, left). The high-sensitivity model achieves a steep induction curve (high apparent Hill coefficient) by using NEQ asymmetric higher-order cooperativities, outperforming alternatives at equilibrium (Fig. 2B, right). These examples demonstrate that NEQ mechanisms can provide functional advantages over equivalent EQ models that utilize the same number of TF binding sites and expression levels (Fig. 2C, D).

Despite the attractiveness of these theoretical ideas, identifying NEQ mechanisms from experimentally measured induction curves can be challenging, as their signatures can be subtle or ambiguous (especially when focusing on individual regulatory phenotypes). Proof-reading could plausibly be detected from the difference in the induction curve plateaus for specific and non-specific TFs (or alternatively from cognate and non-cognate, i.e. mutated, TF binding sites) (Fig. 2B). Detecting higher-order cooperativities necessitates measuring the whole induction curve precisely over a large range of input TF concentrations in a setup where the relevant TF binding sites are fully known. Neither of these tasks appears easy, especially if the predicted effects are not large or if there are substantial systematic uncertainties about the experimental setup.

Recent work has systematically dissected the transcriptional response of a eukaryotic enhancer in the fly embryo to demonstrate the limitations of EQ pairwise cooperativity that is thought to prevail in bacteria, and argue that NEQ higher-order cooperativities between transcription factors and cofactors (such as pioneer factors, Mediator, histone modifiers, etc.) are required to shape the transcriptional response [49]. Beyond steady state induction curves, one could look at the population-level distributions or temporally-resolved measurements. Distributions of mRNA or protein counts can be estimated using fixed approaches (smFISH, immuno-staining). Although full distributions are obviously more informative than their means, they may still be insufficient, however, to definitely distinguish EQ from NEQ mechanisms. Indeed, a steady state distribution can look almost identical (cf. Fig. 1C, the steady state occupancies are identical across all models) or bear very subtle NEQ signatures [26], leading to issues of structural (infinite amount of data cannot discriminate models) or practical identifiability (large amount of data needed to discriminate models).

A better approach is to measure these population distributions along a time course after induction (during relaxation to steady state) [51]. Indeed, for dynamic data already at the level of the means typical signatures of NEQ processes are detectable, e.g. humps in the relaxation curves [32]. Recent work has demonstrated that thermodynamic models of TF binding have difficulties recapitulating the onset and relaxation dynamics for the expression of a eukaryotic gene, while a simple NEQ model of transcription-factor-driven sequential chromatin accessibility accounted for the data better [50]. Even here difficulties abound, mainly because temporal correlations among individual cells are lost in a population measurement, and – as explained in the second section – NEQ signatures are principally exposed through the sequential nature of state transitions in individual systems.

Analyzing temporal fluctuations from single-cell time-series is potentially the most powerful way to discriminate EQ vs NEQ models of regulation [20,21,43,44]. One possible approach would harness auto-correlation functions and waiting time distributions. While auto-correlation functions can be computed directly from gene expression time series using various imaging reporter schemes [52], waiting time distributions necessitate modeling of the time series, typically using Hidden Markov Models, to identify regulatory states of the promoter [21,43]. Recent advances in live imaging enable better identification of such states, by measuring both input TFs and the transcriptional output at defined loci [53,54], or by observing several molecular components (cofactors, Pol2, etc.) simultaneously [55,56]. Thus, there is hope that in the near future NEQ regulatory mechanisms could be inferred directly from measured molecular processes *in vivo*.

Lastly, we stress that certain successes of EQ models in eukaryotic systems cannot be used as evidence for equilibrium processes which would rule out NEQ models. As a pedagogic example, consider energy matrix (or, alternatively, position-weight-matrix (PWM)) models that are often used to describe the sequence-specific occupancy of TFs on DNA [57–59]. Although these models are typically rationalized or derived assuming thermodynamic equilibrium, their predictive success on real data does not necessarily imply that the real system is operating at equilibrium. First, these models are not models of regulation per se but of TF binding or occupancy only; a full regulation model additionally specifies how the information encoded by TF occupancies is integrated into transcriptional output, which could well be via a NEQ mechanism. Our high-specificity toy model (Fig. 2A, left) is a case in point: the TF occupancy is unaffected by the presence or absence of proofreading (in both cases, it is given by $k_+ / (k_+ + k_-)$, independently of the value of $k_q$).

Second, even when energy matrix or PWM models provide great predictive power of expression from sequence assuming functions that are characteristic of equilibrium (e.g., expression is a logistic function of the predicted TF binding energy) [59,60], this does not exclude an equal or even better performance derived under alternative NEQ assumptions. In sum, we should be careful *not* to recognize predictive success of energy matrix models as evidence for EQ (and against NEQ) regulation; even in bacteria, where energy matrix models have been extremely successful, there is some evidence for NEQ processes that could dictate the unbinding of factors [8].

**[Figure 2]** Two toy models of gene regulation with functional advantages when operating out-of-equilibrium (NEQ) versus in equilibrium (EQ). **(A)** (Left) High-specificity model

implements kinetic proof-reading through an irreversible transition with rate $k_q$ leading to the active state (expression E is equal to active state occupancy). The two other states denote TF bound / unbound to the DNA. Ratio $k_-/k_q$ determines the strength of the proof-reading. When $k_q \to \infty$, the model reduces to the simple two state model (unbound and active) at equilibrium. (Right) High-sensitivity model with N = 3 TF binding sites (8 occupancy states) and higher-order cooperativity $a_i^n$ between TFs. Expression E is "all-or-nothing", equal to the occupancy of the all-bound state (active state). Detailed balance holds only when $a_1 = a_2$; the ratio $a_2/a_1$ moves the model out-of-equilibrium by controlling the degree of asymmetry among cooperativities. **(B)** Induction curves. (Left) Induction curves for EQ (blue) and NEQ (orange) proof-reading model with specific TF binding site ($k_- = 1$) and unspecific site ($k_-^{NS} = 10^2$). Both models lead to similar induction curves for specific sites, but NEQ model's expression plateaus at $k_q/(k_- + k_q)$ for non-specific binding whereas the EQ model still plateaus at 1. (Right) Induction for EQ (blue) and NEQ (blue to orange) asymmetric cooperativity models (with same residence time $T_A$). The NEQ models achieve a vast range of sensitivities, defined as the slope at half-maximum expression. **(C)** Optimal operating regime of the two models. (Left) Specificity S, defined as the ratio of expression from specific and non-specific TF binding sites, as a function of the proof-reading ratio $k_-/k_q$, with TF concentration adjusted to hold E = 0.5, shows an optimal regime ($10^{-3} < k_-/k_q < \sim 1$, gray region) where the specificity of a NEQ model clearly outperforms the EQ limit, $k_q \to \infty$. (Right) Hill coefficient H, defined as the log-derivative of the expression curve at half maximum, as a function of the cooperativity ratio $a_2/a_1$, with E = 0.5, and fixed active state residence time, shows an optimal regime ($1 < a_2/a_1 < f(a_1)$, gray region) where the sensitivity of a NEQ model is larger than the EQ limit, $a_1 = a_2$. **(D)** Only a fraction of regulatory phenotype space is accessible at fixed average expression. (Left) All NEQ models outperform the EQ limit (black curve), though the maximal specificity increase is limited. (Right) A fraction of NEQ models outperforms the EQ limit (black curve), with some breaking the limit of H = N = 3 set by Hill-type regulation.

### Normative approach: navigating the space of non-equilibrium regulatory schemes

As the space of possible NEQ gene regulatory models is vast, can we turn to theory for guidance? Here we put forward a normative approach, which assess the functional relevance of regulatory schemes *a priori* [61]. One first identifies phenotypes of regulatory systems, which could have been evolutionarily selected for and which ideally can be measured experimentally. These phenotypes can be of various kinds: static regulatory phenotypes (expression amplitude, specificity, sensitivity), dynamic regulatory phenotypes (noise in gene expression, correlation or relaxation time of the gene expression output), or molecular phenotypes (TF residence time on the DNA, lifetimes of various other molecular complexes, etc.). The key idea is that some of these phenotypes can be experimentally estimated or bounded (e.g., TF residence time on the DNA), while others can be assumed to have been driven by evolution towards their optimal values (e.g., minimizing response time, minimizing noise, maximizing specificity etc.) Together, these assumptions define a constrained optimization problem whose solution identifies not a single model, but rather a class (or ensemble) of models---in terms of their structure and parameter domain---for which we have prior belief that they *may* be simultaneously biologically relevant and consistent with measured constraints and phenomenology [59]. Theory thereby suggests the ensemble to focus our attention and efforts on.

Yet, surely, this approach must strongly depend on our assumptions about the regulatory phenotypes which are being optimized by evolution. We do not hide from this fact. Indeed, the functional relevance of regulatory phenotypes will depend on their biological context:

what is beneficial for a developmental gene might be inappropriate for a housekeeping gene. For instance, the former might have been selected for high sensitivity (sharp spatial gene expression domain boundary during patterning) and short correlation time (fast response) [49,62], whereas the latter may have been selected for low sensitivity and long correlation time (promoting stability with respect to input changes). As with all modeling efforts, assumptions must be made. These assumptions (about evolutionary optimality), however, are no longer vague verbalizations at the end of a research paper, but become embodied in a mathematical formalism that, in the normative approach, we can parametrically vary and statistically assess, given the data. Ultimately, one could hope to identify a single high-level regulatory phenotype that subsumes the others, in order to propose a predictive theory for genetic regulation, paralleling the success of efficient coding in neuroscience [2].

Within the normative approach, one can thus numerically quantify how well a certain (inferred) model achieves a regulatory phenotype – or how far is it from the theoretical optimum. We review such an analysis [44] in Fig. 3, which focuses on a possible model of eukaryotic gene regulation (Fig. 3A), where TFs interact with the Mediator complex to drive expression (Fig. 3B). The normative approach in this context allows us to compare various models quantitatively under identical conditions, for instance, at fixed average expression (cf. Fig. 2D and 3C), or at (experimentally) fixed average TF residence time, and to compare NEQ models to their EQ limit (Fig. 3C). One can discover functions that are inaccessible to EQ models, i.e., regulatory phenotypes that NEQ models can further minimize or maximize, as desired (cf. Fig. 2 for specificity) [24,29]. One can furthermore discover tradeoffs (Fig. 3C): increasing specificity might necessarily lead to higher gene expression noise [48,63]. Lastly, if a single regulatory phenotype is relevant, one can make testable *ab initio* predictions: optimization identifies model parameters which extremize the phenotype given measured constraints and these predictions can, at least in principle, be compared to direct inferences from data.

When the dust settles, the most important upshot of regulatory model space exploration might be the simple observation that the space of NEQ models is vast, and that most of that space is populated by dysfunctional models or models that do not outperform their equilibrium counterparts by any clear measure (Fig. 3D). At first glance, this prospect appears depressing. On the other hand, it means that if evolutionary adaptation *did* act to select for the chosen regulatory phenotypes, our normative approach will rule out most of the parameter space as deleterious, thereby focusing our models and attention into a small sub-space for the parameters and for the model selection (Fig. 3E). From this perspective, such a normative approach should hold great promise for data modeling as well as its functional interpretation, especially when both can be performed within the same formal framework, as recently proposed [59].

**[Figure 3]** Normative approach helps us navigate a complex NEQ gene regulatory model. **(A)** Scheme of the Monod-Wyman-Changeux-like (MWC) model (here with N = 1 TF binding site, simplified from [44]) for a putative eukaryotic enhancer, describing TF un/binding (with rates $k_-$ and $k_+$) and Mediator un/binding (with rates $\kappa_-$ and $\kappa_+$), TF-Mediator interaction (parameter $\alpha$), and a proof-reading step (with rate $k_{link}$) by the formation of a link between the TF and the Mediator. For $k_{link} \to \infty$, this model reduces to classic equilibrium MWC. **(B)** Stochastic realization of the model for N = 3 TF binding sites in EQ regime (blue, $k_{link} \to \infty$) and NEQ regime (orange). (Top) Occupancy of the Mediator-bound (expressing) state as a function of time. (Bottom) Protein counts simulated from the active state of the NEQ model

show bursts due to slow activation dynamics, measured by the noise parameter $\Phi$ [44]. **(C)** Accessible regulatory phenotypes of the model at fixed average expression, such as specificity S, propagated noise $\Phi$ (left colorbar) and sensitivity H (right colorbar), as a function of TF residence time $T_{TF}$. Black line corresponds to the EQ limit. $\Phi$ is in trade-off with S (high specificity implies high noise), while $T_{TF}$ is in trade-off with H (high sensitivity implies high TF residence time). Thus, it is difficult to optimize all regulatory phenotypes at once. The star stands for one possibly relevant model that provides good improvement in S, high H, low $\Phi$ and low $T_{TF}$. **(D)** Varying $k_{link}$ and $\alpha$ enables sampling the regulatory space in (C). The colorbar reflects the magnitude of an arbitrary utility function (cf. [61]) of the combined phenotypes (increasing with S and H, decreasing with $\Phi$ and $T_{TF}$). Gray curves represent equi-phenotype lines. Most parameter values lead to functionally unattractive models (blue region). Only a small subspace (orange region around the star) of models are functionally relevant as they simultaneously optimize multiple phenotypes. **(E)** By first optimizing the regulatory phenotypes, the normative approach can restrict the model or parameter space prior to inference, making model construction and subsequent inference of NEQ models more tractable.

## Future challenges

**Taming the complexity of NEQ models.** Non-equilibrium mechanisms are free of all constraints on their rate parameters that stem from detailed balance. This leads to an unavoidable explosion of free parameters in NEQ models that complicates analysis and inference. One strategy for taming this complexity relies on ongoing progress in non-equilibrium statistical physics. Examples include: better understanding of the minimal energetic cost required to maintain a given NEQ steady state [64]; better understanding of the fluctuations in the different components of NEQ reaction networks and the constraints that these fluctuations must satisfy [65]; better understanding of the symmetries appearing in NEQ reaction network offering prospect of simplification [66]; and finally, development of relevant coarse-graining strategies [15].

Another strategy for taming the complexity is the normative approach that we have advocated here for [61]. As opposed to non-living matter, living systems have been evolutionarily selected for function, which must have resulted in implied selection for – and thus optimization of – various regulatory phenotypes. By identifying models that optimize various phenotypes we essentially restrict (or at least bias) the space of all possible models to a hopefully much smaller sub-space that is functionally relevant prior to further analysis or inference (Fig. 3E). The only technical requirement is the ability to compute the regulatory phenotypes, but this can typically be done, at least numerically, from the master equation. We see the normative approach as complementary, not competing with, other ways to tame the complexity: time will tell whether these exciting theoretical advances actually shed light on real biochemical regulatory networks.

**Energetic costs of regulation.** In eukaryotes, the energetic cost of putative NEQ regulation most likely represents only a small fraction of the total energy budget of a cell, and this cost might well be worth paying for. For instance, NEQ mechanisms (while requiring some energy expenditure) could alleviate even larger energetic costs of spurious or erroneous transcription and translation [24,44] that may be unavoidable for EQ regulatory schemes of lower specificity [65], sensitivity, or sub-optimal response timing [29, 64].

Recent studies in yeast and mammalian cells have demonstrated that transcription and translation, when in excess, can represent a significant burden on global resources [67,68], which cells may try to avoid via optimal resource allocation. Optimal resource allocation arguments thus imply that even if the energy considerations ultimately govern the regulatory architecture, that happens indirectly: not by favoring regulatory schemes that are intrinsically energetically cheap (such as regulation at equilibrium), but rather, by favoring regulatory schemes that minimize the downstream energetic (or, perhaps more broadly, fitness) costs of mis-regulation. Further work is needed to quantify these costs and assess if they strengthen the case for NEQ regulatory schemes.

**Tantalizing cues for NEQ regulation.** Certain reported features of eukaryotic regulation are hard to reconcile with equilibrium mechanisms. It is difficult to understand the high specificity of eukaryotic gene regulation through DNA binding where TFs recognize very short (6-10 bps) motifs, often with individually weak specificity [69,71]. High specificity could emerge from collective effects, either in equilibrium via various cooperative schemes [70,71] or out of equilibrium, e.g., via proof-reading [24,72–74]; it is important to note, however, that high equilibrium cooperativity does not automatically guarantee high specificity [65,71], a fact which often seems overlooked. In addition, the measured residence times of TFs on their specific binding sites tend to be short (a few seconds) and the binding events transient (e.g., as in "hit-and-run" regulation [75]), with the sequential ordering of TF and other cofactor (Mediator, P300, Brd4, etc.) arrivals playing a key role for proper Pol2 initiation and processive elongation [55,56,76,77]. Such a highly transient and sequential assembly of the "enhanceosome", e.g. the pre-initiation complex, can hardly be accounted for by EQ models.

Lastly, the ever-changing chromatin landscape, the sequential accumulation of chromatin marks, and promoter state progression, lead to peaked inactive waiting-time distributions and suggest NEQ regulatory mechanisms [43,44,78]. An important task for the future is therefore to consider a totality of existing and new experimental results integratively, especially those results that have not even been collected with the explicit purpose to test the (non-)equilibrium regulation hypothesis.

## Simulation code for toy-models

The following Matlab functions are provided in a code repository accessible at [INCLUDE REFERENCE]: (i) functions to compute the state rate matrix (Laplacian matrix of the master equation) for each model as a function of input parameters; (ii) a generic function to compute the various regulatory phenotypes and other features (waiting time distributions, fluxes, entropy production, etc.) from the state rate matrix; (iii) a function to generate stochastic realizations of the models; (iv) functions to generate the different figures.

## Acknowledgements

This work was supported through the Center for the Physics of Biological Function (PHY–1734030) and by National Institutes of Health Grants R01GM097275 and U01DK127429 (TG). GT acknowledges the support of the Austrian Science Fund grant FWF P28844 and the Human Frontiers Science Program.

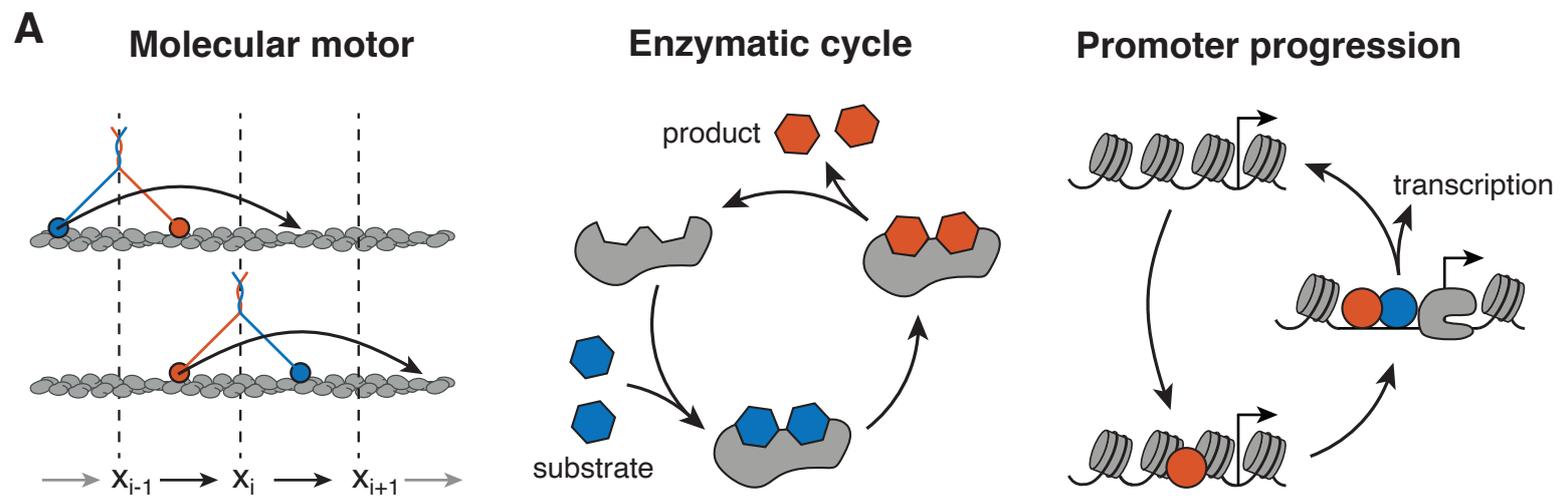

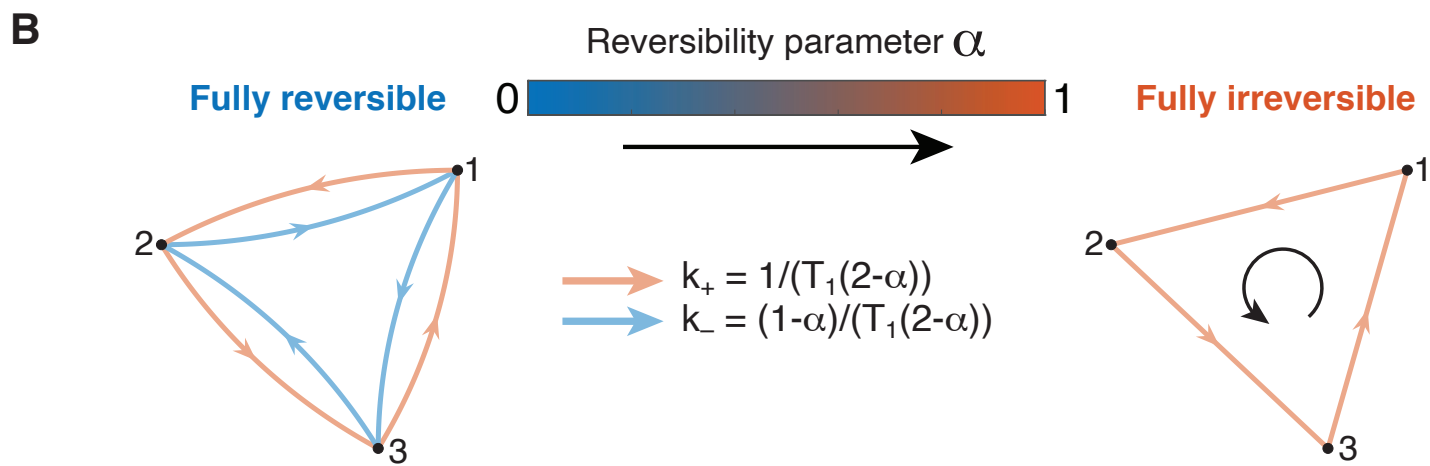

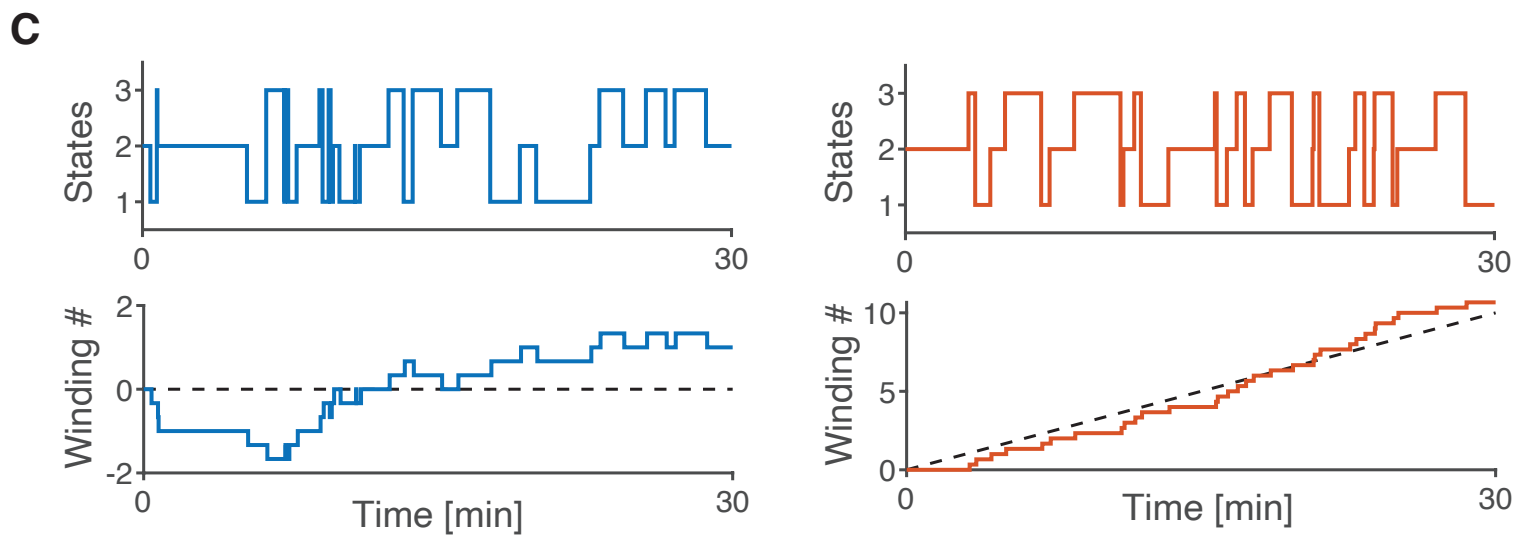

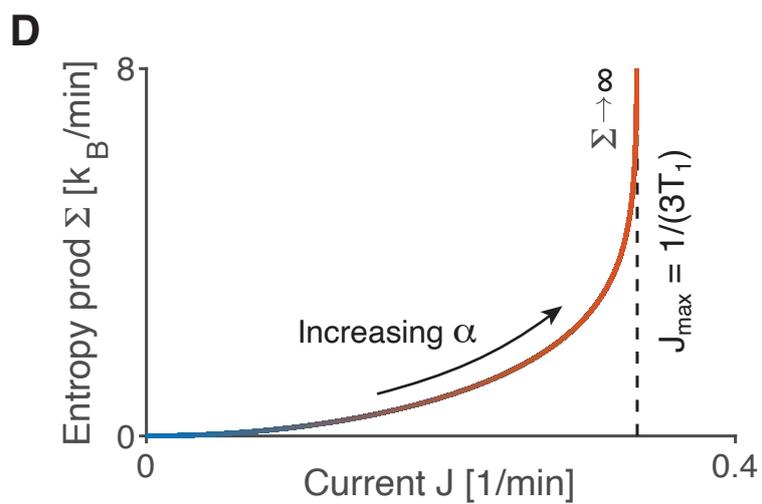
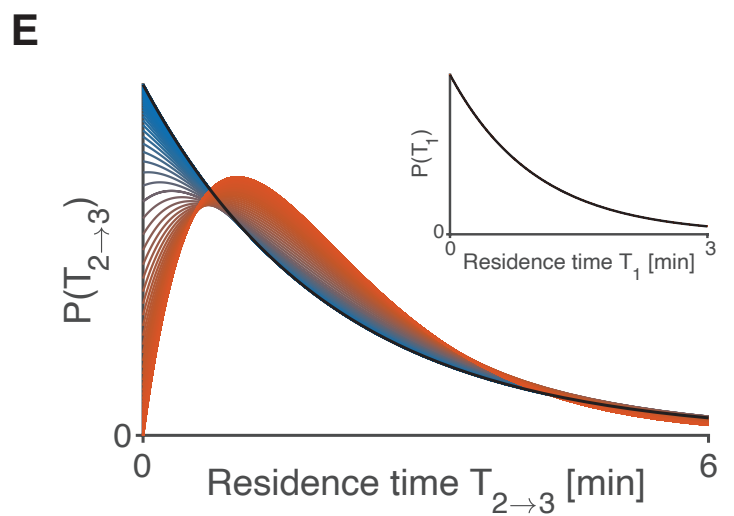

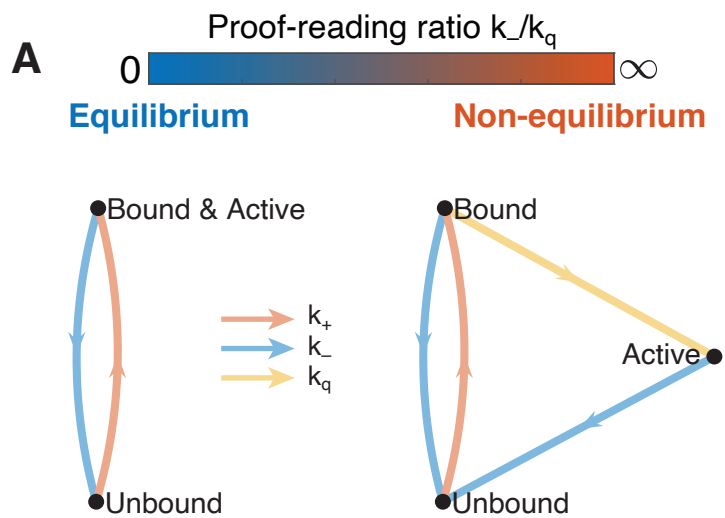
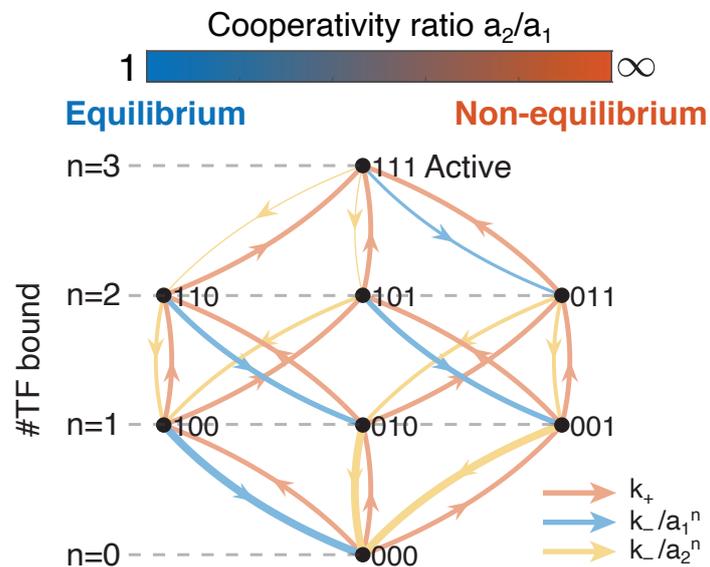
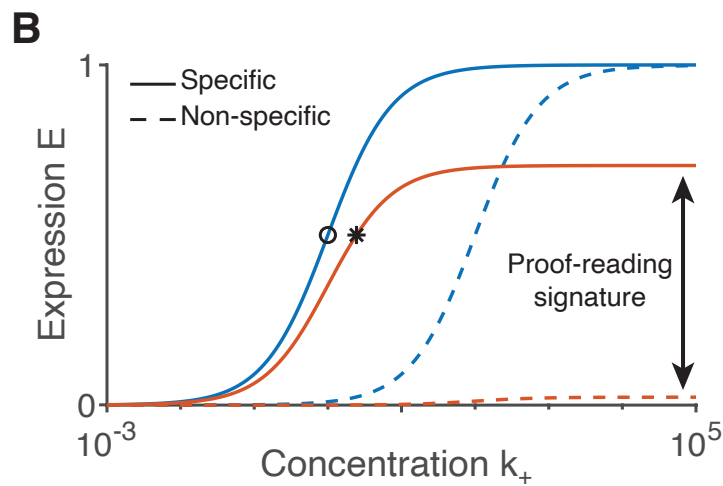
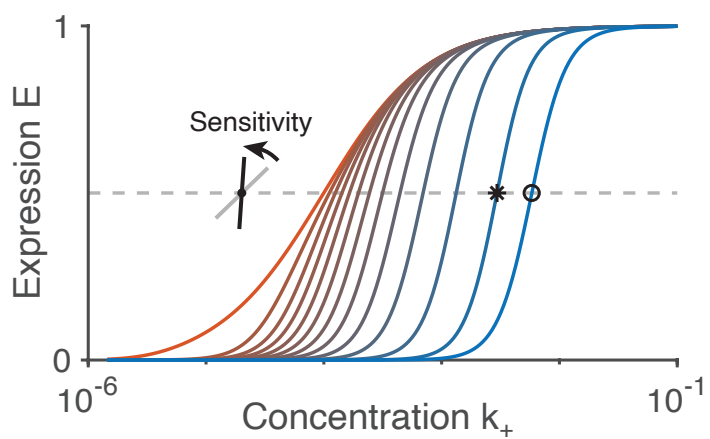
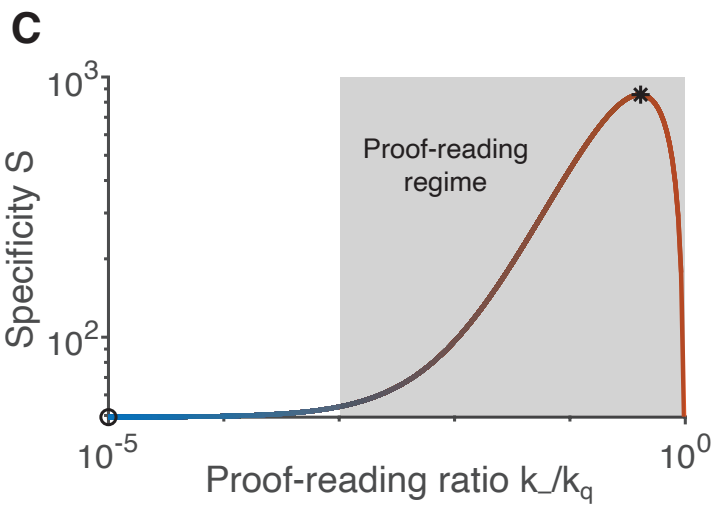
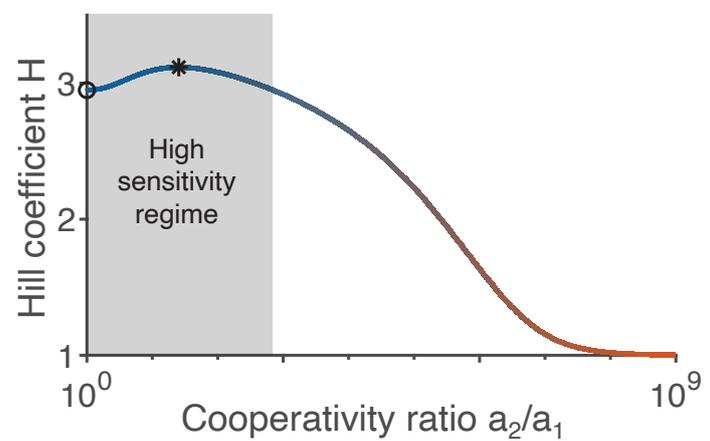
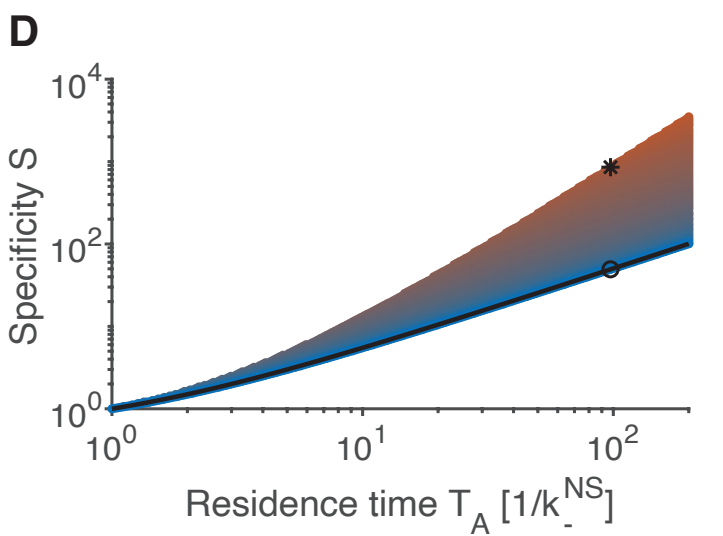
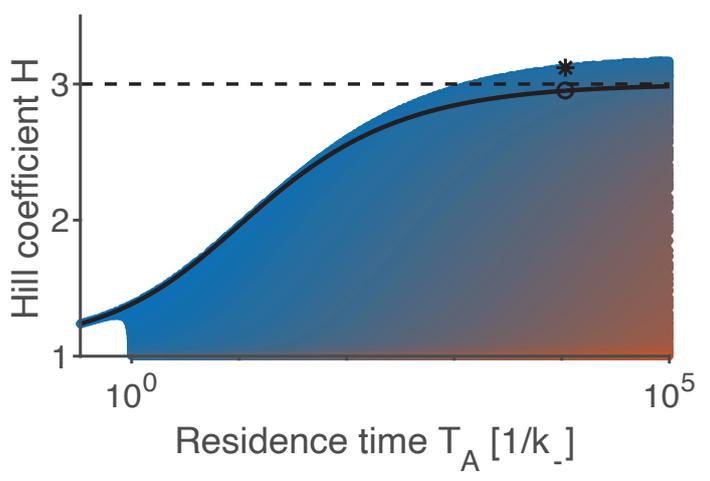

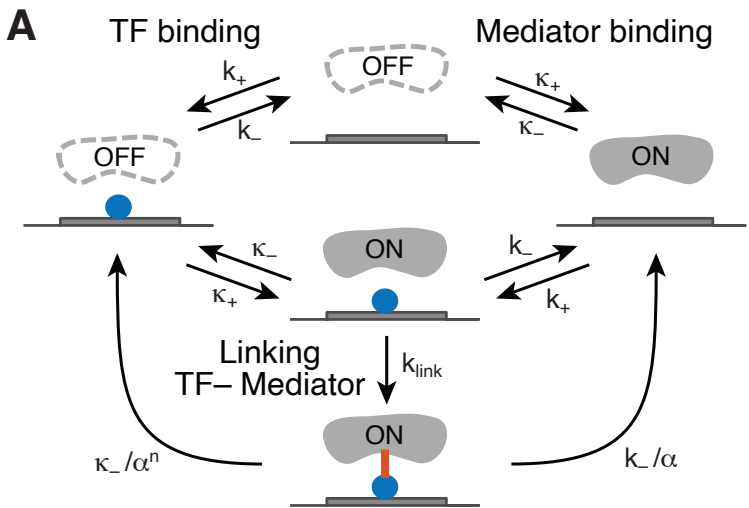
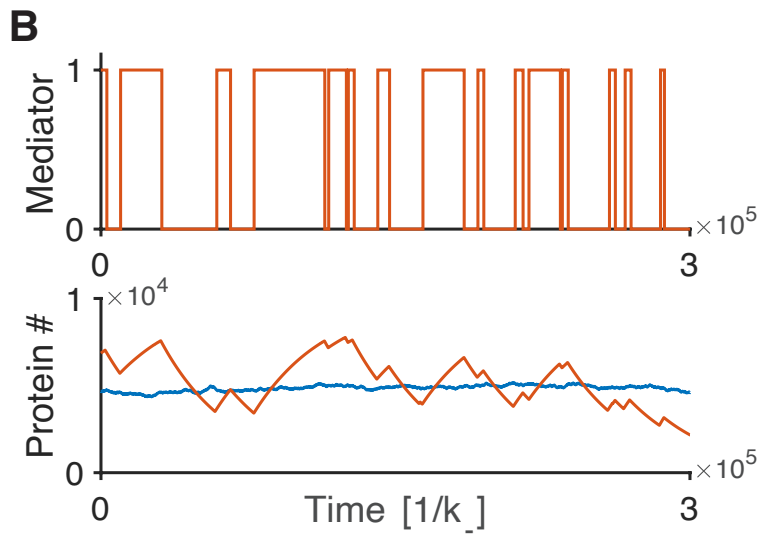
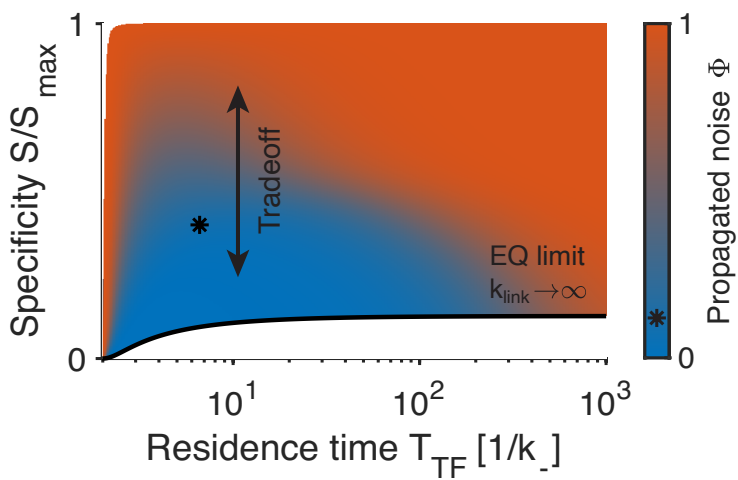
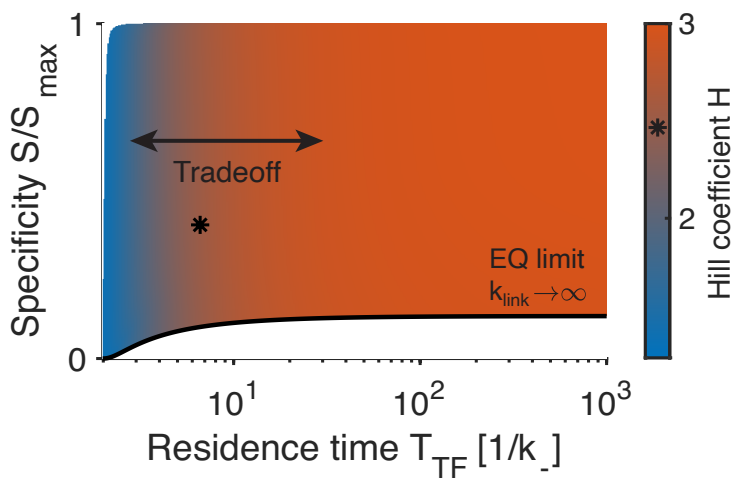
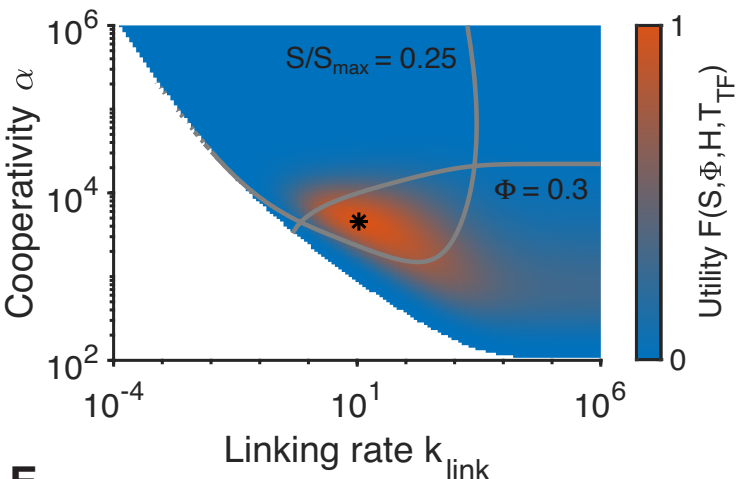
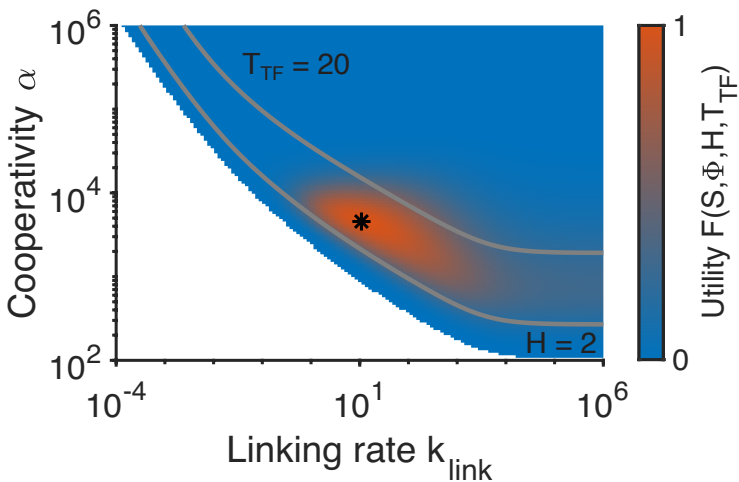
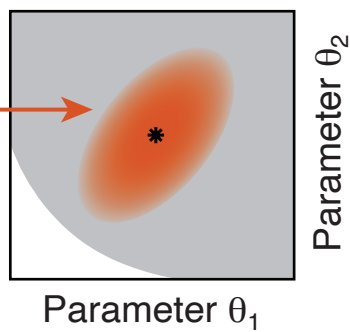
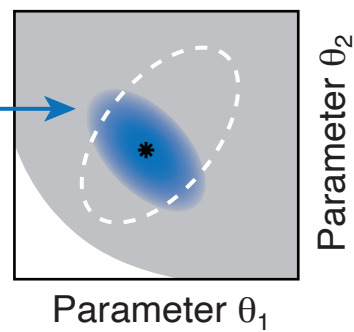